\documentclass[aps,prl,twocolumn,showpacs]{revtex4}
\usepackage{graphicx}

\usepackage{bm}

\newcommand{\ke}[1]{|#1\rangle}

\newcommand{\da}{^\dagger}

\newcommand{\pq}[1]{\left[ #1 \right]}

\newcommand{\av}[1]{\left\langle #1 \right\rangle}

\begin{document}

\title{Cooling trapped atoms in optical resonators}
\author{Stefano Zippilli$^1$}
\author{Giovanna Morigi$^2$}
\affiliation{$^1$ Abteilung f\"ur Quantenphysik,
Universit\"at Ulm, D-89069 Ulm, Germany\\
$^2$ Departament de Fisica, Universitat Aut\`onoma de Barcelona, 08193 Bellaterra, Spain}

\date{\today}
\begin{abstract}
We derive an equation for the cooling dynamics of the quantum
motion of an atom trapped by an external potential inside an
optical resonator. This equation has broad validity and allows us to
identify novel regimes where the motion can be efficiently cooled
to the potential ground state. Our result shows that the motion is
critically affected by quantum correlations induced by the
mechanical coupling with the resonator, which may lead to
selective suppression of certain transitions for the appropriate
parameters regimes, thereby increasing the cooling efficiency.
\end{abstract}
\maketitle

Cavity cooling is a recent expression, which stresses the role of
the mechanical effects of a resonator on the atomic and molecular
center-of-mass dynamics. Indeed, the coupling between resonator
and atom gives rise to complex dynamics, one aspect of which is
the substantial modification of the atom spectroscopic
properties~\cite{CQED94}. This property allows one to change the
atom scattering cross section, thereby affecting, and eventually
tailor, the mechanical dynamics of the atomic center of
mass~\cite{Domokos03}.
Moreover, the motion of the atom 
changes the medium density, thereby affecting the resonator field itself. 
Several recent experiments have reported relevant features of these complex
dynamics~\cite{Rempe04,KimbleFORT03,Vuletic03,Ring03,Zimmerman04,Buschev04}, 
Experimental demonstrations of atom cooling in
resonators~\cite{Rempe04,KimbleFORT03,Vuletic03,Ring03,Zimmerman04}
have shown, among others, that cavities are a promising tool for
preparing and controlling cold atomic samples of scalable
dimensions, which may find relevant applications, for instance in quantum
information processing~\cite{QuInfo}.

\begin{figure}[ht]
\includegraphics[width=8cm]{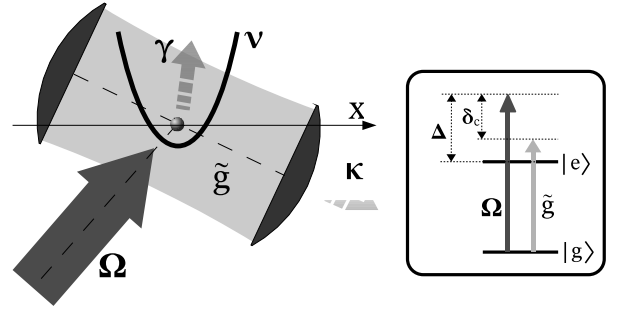}
\caption{An atom is confined by an external potential inside an
optical resonator, and couples to a laser and a mode of the
resonator. Energy is dissipated by spontaneous emission, at rate
$\gamma$, and by cavity decay, at rate $\kappa$. Inset: The dipole
levels $\ke{g}$ and $\ke{e}$. $\Delta$ and $\delta_c$ are the
detunings of dipole and cavity from the laser frequency.}
\label{Fig:1L}
\end{figure}

In this letter we present a study of the quantum dynamics of the
center of mass motion of an atomic dipole, which couples to a
resonator and to a laser field driving it from the side. The
system is sketched in Fig.~\ref{Fig:1L}. Differently from recent theoretical works
on cavity cooling of atomic clouds~\cite{Domokos02}, here the center of mass is
confined by a tight trap, in a configuration that may correspond
to the experimental situations reported for instance
in~\cite{KimbleFORT03,IonCQED,Sauer03}. Starting from a master
equation for the quantum variables of dipole, cavity and
center-of-mass, we derive a closed equation for the center-of-mass
dynamics. This equation generalizes previous theoretical
studies~\cite{Cirac95,Vuletic01} and allows us to identify novel
parameter regimes, where cooling can be efficient. Moreover, its
form permits us to identify the individual scattering contributions, thereby gaining insight into the role of the various physical parameters. We show that
quantum correlations between atom and resonator 
may lead to the suppression of scattering transitions, thereby
enhancing the cooling efficiency.

The starting point is the master equation for an atom of mass $M$
whose dipole transition between the ground and excited states
$\ke{g}$ and $\ke{e}$ couples (quasi-)resonantly to a laser and
the mode of an optical resonator with wave vector $k_L$ and $k_c$,
respectively ($|k_L|\approx |k_c|=k$). In the reference frame of
the laser the state $\ke{e}$ is at frequency $-\Delta$ and the
cavity mode at~$-\delta_c$. The atomic center-of-mass is trapped by a
harmonic oscillator of frequency $\nu$ along the $x$-axis, while
the motion in the orthogonal plane is frozen out. Later on we discuss how the treatment can be generalized to three dimensional motion. The density matrix $\rho$ of
cavity mode and atom internal and external degrees of freedom
evolves according to the master equation $\partial\rho/\partial
t={\cal L}\rho$ where the Liouvillian ${\cal L}$ describes the
coherent and incoherent dynamics. External and internal degrees of
freedom are coupled through the spatial gradient of the
interaction with the electromagnetic field, which scales with the
Lamb-Dicke parameter $\eta=\sqrt{\hbar
k^2/2M\nu}$~\cite{Stenholm86}. We assume tight confinement, such
that the gradient of the e.m.-field over the atomic wave packet is
small and the Lamb-Dicke regime holds, $\eta \ll 1$, and treat the
coupling of the center-of-mass with dipole and cavity variables in
perturbation theory. At second order in $\eta$ we decompose ${\cal
L}={\cal L}_0+\eta {\cal L}_1+\eta^2{\cal L}_2+{\rm o}(\eta^3)$
where the subscripts indicate the corresponding order in the
expansion. The term ${\cal L}_0={\cal L}_{0E}+{\cal L}_{0I}$ is
the sum of the Liouvillian for the external ($E$) and internal and
cavity degrees of freedom ($I$). Here, ${\cal L}_{0E}\rho=[H_{\rm
mec},\rho]/{\rm i}\hbar$ and $H_{\rm mec}=\hbar\nu(\hat{n}+1/2)$
center-of-mass harmonic oscillator with $\hat{n}$ number operator
for its phononic excitations; the term ${\cal
L}_{0I}\rho=[H_0,\rho]/{\rm i}\hbar+({\cal L}_s+{\cal K})\rho$,
where ${\cal K}$ and ${\cal L}_s$ describe cavity decay at rate
$\kappa$ and dipole spontaneous emission at rate $\gamma$,
respectively~\cite{Carmichael93}, and
$H_0=-\hbar\Delta\sigma\da\sigma-\hbar\delta_ca\da
a+\hbar[\sigma\da(\Omega+\tilde{g} a)+{\rm H.c.}]$, with
$\sigma=\ke{g}\langle e|$, $a,a\da$ annihilation and creation operators
of a
cavity photon, $\tilde{g}$ coupling constant between cavity and
dipole at the center of the trap, and $\Omega$ laser Rabi
frequency. The first-order correction ${\cal
L}_1\rho=[V,\rho]/{\rm i}\hbar$ contains the
operator $V=(V_L+ V_c)(b+b\da)$, which gives the mechanical coupling
induced by laser and resonator, whereby $b$,  $b\da$ are the creation and annihilation operators of a vibrational quantum, the operators 
\begin{eqnarray}\label{1stlaser}
V_L&=&{\rm i}\hbar\varphi_L \Omega(\sigma\da-\sigma)\\
V_c&=&\hbar \varphi_c\tilde{g}(a\sigma\da+a\da
\sigma)\label{1stcavity}
\end{eqnarray}
and the coefficients $\varphi_L$, $\varphi_c$ scale the
corresponding recoil and depend on the geometry of the
setup~\cite{Footnote1}. 
The term ${\cal L}_2$ describes mechanical
effects associated with the spontaneous emission of a photon,
which determine diffusion, and state-dependent energy
shifts~\cite{Javanainen84,Cirac92}. The modifications induced by
these shifts in the cavity field dynamics have been discussed
in~\cite{Rice04}. In the present treatment, where $\eta\ll 1$,
they are high-order corrections to the external potential and
hence do not appreciably affect the center-of-mass dynamics.

At second order in the Lamb-Dicke expansion we derive
a rate equation for the occupations of the number states $\ke{n}$ of
the center-of-mass harmonic oscillator~\cite{Javanainen84,ZippilliNew}.
The rate equation is expressed in terms of the heating rate $\Gamma_{n\to n+1}=\eta^2(n+1)A_+$ and the cooling rate $\Gamma_{n\to n-1}=\eta^2 nA_-$, which describe transitions that change $\ke{n}$ by one excitation. Here,
\begin{eqnarray*}
A_{\pm}={\rm Re}\{2{\rm Tr}_I\{V({\cal L}_{0I}+{\rm
i}\nu)^{-1}V\rho_0\}+\alpha\gamma {\rm Tr}\{ |e\rangle\langle
e|\rho_0\}\}
\end{eqnarray*}
with ${\rm Tr}_I$ trace over the dipole and cavity degrees of
freedom and $\alpha$ geometric factor giving the angular
dispersion of atomic momentum due to the random recoil of spontaneous
emission. Their explicit evaluation requires the knowledge of the
spectrum of the correlation of the operator $V$ and of the steady
state $\rho_0$ of dipole and cavity
mode at zero order in~$\eta$~\cite{Javanainen84,Cirac92,Morigi03}. When the laser weakly perturbs the atom-cavity ground state $\ke{\phi_0}=\ke{g,0_c}$, we find~\cite{ZippilliNew}
\begin{eqnarray}\label{Apm}
A_{\pm}
&=&\gamma\alpha|{\cal T}_S|^2 + \gamma|\varphi_L{\cal T}_L^{\gamma,\pm} + \varphi_c{\cal T}_c^{\gamma,\pm}|^2\\
& &+\kappa|\varphi_L{\cal T}_L^{\kappa,\pm}+\varphi_c{\cal
T}_c^{\kappa,\pm}|^2 \nonumber
\end{eqnarray}
where
\begin{eqnarray}\label{T:S}
&&{\cal T}_S=\Omega\frac{\delta_c+{\rm i}\kappa/2}{f(0)}\\
&&{\cal T}_L^{\gamma,\pm}={\rm i}\Omega\frac{(\delta_c\mp\nu+{\rm
i}\kappa/2)}{f(\mp\nu)}
\label{T:L:G}\\
&&{\cal T}_L^{\kappa,\pm}={\rm i}\Omega\frac{\tilde g}{f(\mp\nu)}\label{T:L:K}\\
&&{\cal T}_c^{\gamma,\pm}=-\Omega\frac{\tilde g^2(2\delta_c\mp\nu+{\rm i}\kappa)}{f(0)f(\mp\nu)}\label{T:C:G}\\
&&{\cal T}_c^{\kappa,\pm}=-\Omega\frac{\tilde
g\pq{(\Delta\mp\nu+{\rm i}\gamma/2)(\delta_c+{\rm
i}\kappa/2)+\tilde g^2}}{f(0)f(\mp\nu)}\label{T:C:K}
\end{eqnarray}
with
\begin{eqnarray}\label{transitions:F}
f(x)=(x+\delta_c+{\rm i}\kappa/2)(x+\Delta+{\rm i}\gamma/2)-\tilde
g^2
\end{eqnarray}
Equations~(\ref{Apm}) and~(\ref{T:S})-(\ref{T:C:K}) are the main
result of this letter. They have been derived for one-dimensional
motion. However, since at second order in $\eta$ the three
directions of oscillation decouple in an anisotropic trap, they can be generalized to
three-dimensional motion as they hold for any geometry of the
setup. Their analytic form allows one for insight into these
complex dynamics, which we now discuss.

\begin{figure*}[t]
\includegraphics[width=18cm]{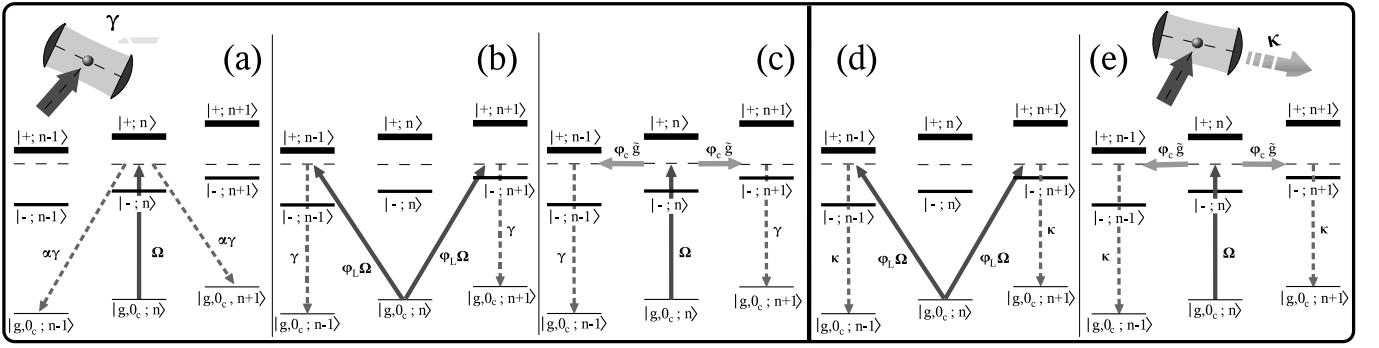}
\caption{Scattering processes leading to a change of the
vibrational number by one phonon. The states $|g,0_c;n\rangle$,
$\ke{\pm;n}$ are the cavity-atom dressed states at phonon number
$n$. Processes (a)--(c) describe scattering of a laser photon
by spontaneous emission. They prevail in good resonators, for
$\kappa\ll\tilde g,\gamma$. Here, (a) ${\cal T}_S$; (b) ${\cal
T}_L^{\gamma,\pm}$; (c) ${\cal T}_c^{\gamma,\pm}$. Processes (d)
and~(e) describe scattering of a laser photon by cavity decay.
They prevail in bad resonators, for $\gamma\ll\tilde g,\kappa$.
Here, (d) ${\cal T}_L^{\kappa,\pm}$; (e) ${\cal
T}_c^{\kappa,\pm}$. See text.}\label{Fig:2L}
\end{figure*}

The rates are the incoherent sum of three contributions. The first
term on the right-hand side (rhs) of Eq.~(\ref{Apm}), $\gamma\alpha|{\cal T}_S|^2$,
describes absorption of a laser photon and spontaneous emission,
whereby the change in the center-of-mass state is due to the
recoil induced by the spontaneously emitted photon. This process
is shown in Fig.~\ref{Fig:2L}(a), where the excitation paths are
depicted in the dressed states basis $\ke{\pm;n}$ of cavity and
atom, which are superposition of the states $\ke{g,1_c}$ and
$\ke{e,0_c}$~\cite{Domokos03} at fixed number of phonons $\ke{n}$.
The scattering rate is scaled by the geometric factor $\alpha$ and
is found after averaging over the solid angle of photon emission
into free space. This contribution is diffusive, as the motion can
be scattered into a higher or lower vibrational state with
probabilities depending on the overlap integrals~\cite{Cirac92}.
Diffusion can be suppressed, namely term~(\ref{T:S}) can
become very small, in good resonators, $\kappa\ll \tilde{g}$, and
for $\delta_c=0$. In this limit the corresponding dynamics are
characterized by destructive interference between the two
excitation paths
$\ke{\phi_0;n}\to\ke{\pm;n}$~\cite{Alsing92,Zippilli04,Rice96}.
This situation appears at zero order in the mechanical effects and
is reminiscent of analogous phenomena encountered in the cooling
dynamics of trapped multi-level atoms~\cite{Morigi00}.

The second term on the rhs of Eq.~(\ref{Apm}),
$\gamma|\varphi_L{\cal T}_L^{ \gamma,\pm }+\varphi_c{\cal
T}_c^{\gamma,\pm }|^2$, describes scattering of a laser photon by
spontaneous emission where the motion is changed by mechanical
coupling to the laser (${\cal T}_L^{ \gamma,\pm }$) and to the
cavity (${\cal T}_c^{\gamma,\pm }$) field. Process~${\cal T}_L^{
\gamma,\pm }$ scales with the geometric factor $\varphi_L$, which
accounts for the recoil due to absorption of a laser photon, and
is depicted in Fig.~\ref{Fig:2L}(b). The transition
amplitude~${\cal T}_c^{ \gamma,\pm }$, shown in
Fig.~\ref{Fig:2L}(c), describes the mechanical coupling between
the dressed states $\ke{\pm}$ due to the resonator. It thus scales
with the geometric factor $\varphi_c$ which accounts for the
recoil due to interaction with the cavity mode. Since the final
state of the two scattering processes is the same, they
interfere~\cite{Footnote2}. In addition, each term is composed by
multiple excitation paths, and can vanish in some parameter
regimes. A representative case is found in good resonators,
$\kappa\ll\tilde g$, when the laser is orthogonal to the motion
axis ($\varphi_L=0$). Then, the transition amplitude, ${\cal
T}_c^{\gamma,+}$ is suppressed for $\delta_c=\nu/2$, as one can
see in Eq.~(\ref{T:C:G}). This suppression arises from destructive
interference between the multiple paths of mechanical coupling via
the dressed states $\ke{\pm;n}$ and $\ke{\pm;n+1}$, thereby giving
a vanishing heating rate. Similarly, the cooling transition can be
suppressed by choosing $\delta_c=-\nu/2$.

The third term in the rhs of Eq.~(\ref{Apm}),
$\kappa|\varphi_L{\cal T}_L^{\kappa,\pm }+\varphi_c{\cal
T}_c^{\kappa,\pm}|^2$, describes scattering of a laser photon by
cavity decay, where the motion is changed by mechanical coupling
to the laser (${\cal T}_L^{\kappa,\pm}$) and to the cavity (${\cal
T}_c^{\kappa,\pm}$) field. The scattered photon is transmitted
through the cavity mirrors into the external modes, and therefore
these two processes do not interfere with the ones above
discussed. They are depicted in Fig.~\ref{Fig:2L}(d) and~(e) and
add up coherently with one another. An interesting situation has
been discussed in~\cite{Cirac95}, where it was shown that
interference between these two terms can lead to suppression of
the heating transition in bad resonators. The result
of~\cite{Cirac95} is recovered from Eqs.~(\ref{Apm})-(\ref{T:C:K})
in the corresponding parameter regime~\cite{ZippilliNew}.

\begin{figure}[!h]
\includegraphics[width=9cm]{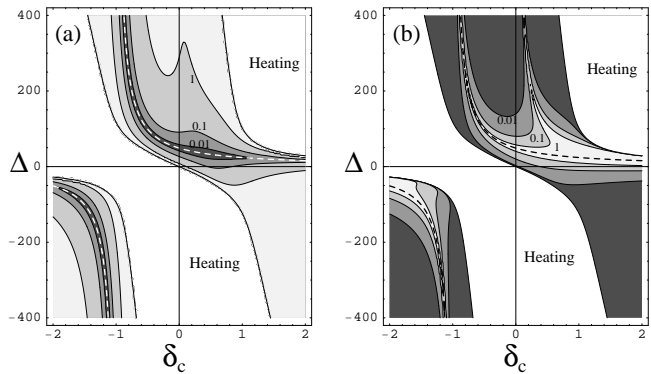}
\caption{Contour plot of (a) the average phonon number at steady
state $\langle n\rangle_{\infty}$ and (b) the corresponding
cooling rate $W$ (units of $10^{-3}\nu$) as a function
of $\delta_c$ and $\Delta$ (units of $\nu$) in the good cavity
limit. The darkest region 
corresponds to the smallest values, the lightest region to the largest values.
The numbers indicate the values at the corresponding contour line. The heating regions are not coded and explicitly indicated. The dashed line represents Eq.~(\ref{Delta}). 
Here, $\eta=0.1$, $\varphi_L=\varphi_c=\sqrt{1/2}$, $\Omega=\nu$,
$\tilde g=7\nu$, $\gamma=10\nu$, $\kappa=0.01\nu$.} \label{Fig:3L}
\end{figure}

The general dynamics are a competition of all these processes. By
inspection of the form of Eqs.~(\ref{Apm}-\ref{transitions:F}),
one can identify a strategy for obtaining efficient
cooling by searching for the parameters such that
$\mbox{Re}\{f(\nu)\}=0,$ thereby minimizing the denominator of
$A_-$, and therefore enhancing the cooling rate over the heating
rate. This condition leads to the equation
\begin{eqnarray}\label{Delta}
\Delta_{\rm opt}(\delta_c)\equiv \frac{\tilde
g^2+\gamma\kappa/4}{\delta_c+\nu}-\nu
\end{eqnarray}
which relates the cavity detuning $\delta_c$ to the atom detuning
$\Delta$ for fixed couplings and decay rates, such that, when
$\Delta=\Delta_{\rm opt}(\delta_c)$, the cooling rate $A_-$ is
maximum. In general, Eq.~(\ref{Delta}) sets the best parameters
for ground state cooling, provided that either
$\gamma\ll\nu$ or $\kappa\ll\nu$. Below we discuss cooling in the
good cavity limit. The bad cavity limit will be discussed
elsewhere. Figures~\ref{Fig:3L}(a) and~(b) are
contour plots, displaying the average number of phonon at steady
state, $\av{n}_{\infty}=A_+/(A_- - A_+)$, and the cooling rate,
$W=\eta^2 (A_- - A_+)$, as a function of $\Delta$ and $\delta_c$
for a good cavity and $\kappa\ll\nu\ll\gamma$, namely when
processes in Fig.~\ref{Fig:2L}(a)-(c) are dominant. These curves
have been obtained from Eq.~(\ref{Apm}), and their validity has
been checked by comparison with full numerical
simulations~\cite{ZippilliNew}. The dashed line indicates the
curve $\Delta=\Delta_{\rm opt}(\delta_c)$. It is evident that this
curve corresponds to the parameter region where the lowest
temperature is achieved. The final limit is set by the ratio
$\kappa/\nu$, since in this regime $\kappa$ is the narrowest linewidth one can achieve for the scattering
processes~\cite{Footnote3}.

Good parameter regimes are found for relatively small values of
$\Delta$ and $\delta_c$, in the range $0<\delta_c<\nu$,
$\Delta>0$, where one has low temperatures and relatively large
cooling rates for a wide range of values. These regions correspond
to novel cooling dynamics, which are characterized by suppression of
heating and diffusive transitions by quantum correlations induced
by the resonator. Here, for instance, the potential ground-state
occupation can reach $99\%$ in a time $\tau\sim 2$~msec  for an
atom in a trap with frequency $\nu=2\pi\times 500$ KHz and $\eta=0.1$, 
whose dipole has line width $\gamma=2\pi\times 5$ MHz and couple to a
resonator with $g=2\pi\times 5$ MHz and $\kappa=2\pi\times 100$
KHz. Larger values of $\kappa$ lead to smaller
occupations, which scale like $\langle
n\rangle_{\infty}\sim\kappa^2/\nu^2$ in good resonators for
$\kappa<\nu$. Figure~\ref{Fig:3L} shows also that the dynamics are
relatively insensitive to fluctuations of the parameters in the
range $0<\delta_c<\nu$, $\Delta>0$. This implies that several
oscillation modes, like the ones of an ion chain~\cite{Morigi04},
could be simultaneously cooled. We remark that good cooling
regions are found in Fig.~\ref{Fig:3L} also for large $|\Delta|$
and $\delta_c\sim -\nu$. This is the so--called cavity sideband
cooling regime, discussed in~\cite{Vuletic01}. Here, enhancement
of the cooling rate is achieved by means of resonance
with the very narrow transition, at large
$|\Delta|$, between the dressed states 
of the cavity-atom coupling. This regime exhibits a critical sensitivity to
variations of the value of $\delta_c$ with respect to $\Delta$.

To conclude, we have presented the explicit form of 
cooling and heating rates of the center-of-mass motion of an 
atom, trapped in an optical
resonator and driven by a laser. This result permits us to
identify the basic processes determining the center-of-mass
dynamics and the parameter regions where cooling can be
efficiently obtained. Novel regimes have been identified, where
the motion is cooled by exploiting destructive interference in the
multiple paths of the mechanical excitations. The resulting
interference bases itself on the discreteness of the vibrational
spectrum, which is the same for the dipolar ground and the excited
state. Some analogies with interference processes in cooling of
trapped multilevel atoms~\cite{Marzoli94,Morigi00} are due to a
similar dressed state structure at low excitation, nevertheless
the mechanical effects induced by the resonator give rise to a
wider wealth of phenomena. Dynamics will be substantially modified
when the external potential depends appreciably on the internal
state~\cite{Rice04,Morigi03}. In future works we will investigate how the dynamics are
modified by collective effects, when many atoms are trapped inside
the cavity.

Beyond applications to cooling, these results allow one for
gaining insight in the complex dynamics of the mechanical effects
of optical resonators on atoms, and their wealth of phenomena
could be exploited for implementing coherent control of this kind of
systems. 

We thank Helmut Ritsch and Axel
Kuhn for discussions. Support from the IST-network QGATES and the spanish Ministerio de
Educaci\'on y Ciencia (Ramon-y-Cajal fellowship 129170) are acknowledged.


\begin{thebibliography}{99}
\bibitem{CQED94}
see for instance {\it Cavity Quantum Electrodynamics}, ed.\ by
P.R.\ Berman, Academic Press (New York, 1994).
\bibitem{Domokos03}
P.\ Domokos, H.\ Ritsch, J.\ Opt.\ Soc.\ Am.\ B {\bf 20}, 1098
(2003).
\bibitem{Rempe04}
P.\ Maunz, {\it et al}, Nature {\bf 428}, 50 (2004).
\bibitem{KimbleFORT03}
J.\ McKeever {\it et al}, Phys.\ Rev.\ Lett.\ {\bf 90}, 133602
(2003).
\bibitem{Vuletic03}
H.W.\ Chan {\it et al}, Phys.\ Rev.\ Lett.\ {\bf 90}, 063003
(2003); A.T.\ Black {\it et al}, Phys.\ Rev.\ Lett.\ {\bf 91},
203001 (2003).
\bibitem{Ring03}
D.\ Kruse {\it et al}, Phys.\ Rev.\ Lett.\ {\bf 91}, 183601
(2003); B.\ Nagorny {\it et al}, Phys.\ Rev.\ Lett.\ {\bf 91},
153003 (2003).
\bibitem{Zimmerman04}
C. von Cube, {\it et al}, Phys.\ Rev.\ Lett.\ {\bf 93}, 083601 (2004).
\bibitem{Buschev04}
P.\ Bushev, {\it et al}, Phys.\ Rev.\ Lett.\ {\bf 92}, 223602
(2004).
\bibitem{QuInfo}
See for instance, {\it The Physics of Quantum Information}, ed.\
by D.\ Bouwmeester, A.K.\ Ekert, A. Zeilinger, Springer-Verlag
(Berlin, Heidelberg, New York, 2000).
\bibitem{Domokos02}
P.\ Domokos and H.\ Ritsch, Phys.\ Rev.\ Lett.\ {\bf 89}, 253003
(2002); 
P.\ Domokos, Th.\ Salzburger, H.\ Ritsch,
Phys.\ Rev.\ A {\bf 66}, 043406 (2002);
P.~Domokos, A.~Vukics, H.~Ritsch, Phys.\ Rev.\ Lett.\ {\bf 92}, 103601 (2004).
\bibitem{IonCQED} M.\ Keller,
{\it et al}, Nature {\bf 414}, 49 (2004); A.B.\ Mundt, {\it et
al}, Phys.\ Rev.\ Lett.\ {\bf 89}, 103001 (2002).
\bibitem{Sauer03}
J.A.\ Sauer {\it et al}, Phys.\ Rev.\ A {\bf 69}, 051804 (2004).
\bibitem{Cirac95}
J.I.\ Cirac, M.\ Lewenstein, P.\ Zoller, Phys.\ Rev.\ A {\bf 51},
1650 (1995).
\bibitem{Vuletic01}
V.~Vuletic, H.W.~Chan, A.T.~Black, Phys.\ Rev.\ A {\bf 64}, 033405
(2001).
\bibitem{Stenholm86} S.\ Stenholm, Rev.\ Mod.\ Phys.\ {\bf
58}, 699 (1986).
\bibitem{Carmichael93}
H.J.\ Carmichael, {\it An Open Systems Approach to Quantum
Optics}, Springer-Verlag (Berlin, Heidelberg, New York, 1993).
\bibitem{Footnote1}
For instance,
$\varphi_L=\cos\theta$ when the laser is a traveling wave forming
the angle $\theta$ with the axis of the motion.
\bibitem{Javanainen84}
J.~Javanainen, M.~Lindberg, S.~Stenholm, J.\ Opt.\ Soc.\ Am.\ B
{\bf 1}, 111 (1984).
\bibitem{Cirac92}
J.I.\ Cirac, {\it et al}, Phys.\ Rev.\ A {\bf 46}, 2668 (1992).
\bibitem{Rice04}
J.\ Leach, P.R.\ Rice, Phys.\ Rev.\ Lett.\ {\bf 93}, 103601
(2004).
\bibitem{ZippilliNew}
S.~Zippilli, G.~Morigi, quant-ph/0508075.
\bibitem{Morigi03} G.\ Morigi, Phys.\ Rev.\ A {\bf 67},
033402 (2003).
\bibitem{Alsing92}
P.M.\ Alsing, D.A.\ Cardimona, H.J.\ Carmichael, Phys.\ Rev.\ A
{\bf 45}, 1793 (1992).
\bibitem{Zippilli04}
S.\ Zippilli, G.\ Morigi, H.~Ritsch, Phys. Rev. Lett. {\bf 93},
123002 (2004); Eur.\ Phys.\ J.\ D  {\bf 31}, 507 (2004).
\bibitem{Rice96}
P.R.\ Rice, R.J.\ Brecha, Opt.\ Comm.\ {\bf 126}, 230 (1996).
\bibitem{Morigi00}
G.\ Morigi, J.\ Eschner, C.H.\ Keitel, Phys.\ Rev.\ Lett. {\bf
85}, 4458 (2000); J.\ Evers, C.H.\ Keitel, Europhys.\ Lett.\ {\bf
68}, 370 (2004)
\bibitem{Footnote2}
They also interfere with the amplitude ${\cal T}_S$. This
interference depends on the angle of emission in free space and is
thus averaged out after integrating over the solid angle.
\bibitem{Footnote3}
Lowest occupations can be reached by using a standing wave
drive~\cite{ZippilliNew}.
\bibitem{Morigi04}
G.~Morigi, Sh.~Fishman, Phys.\ Rev.\ Lett.\ {\bf 93}, 170602
(2004).
\bibitem{Marzoli94}
I.\ Marzoli, {\it et al}, Phys.\ Rev.\ A {\bf 49}, 2771 (1994).
\end{thebibliography}
\end{document}